\documentclass[12pt]{article}
\usepackage{amsfonts, amsmath, amssymb, cite, graphicx, }

\usepackage{color}
\usepackage[all, knot]{xy}
\usepackage{tikz}
\usepackage{subfigure}
\usepackage{braket}

\usepackage[utf8]{inputenc}
\usepackage{ulem}

\usepackage{bm}
\usepackage{epstopdf}
\usepackage[footnotesize]{caption}
\usepackage{amsthm}
\usepackage{amsthm,amsmath,amssymb}
\usepackage{enumitem}
\usepackage{mathrsfs}
\usepackage{makecell}
\usepackage{diagbox}

\usepackage[margin=3cm]{geometry}

\begin{document}

\begin{titlepage}
\vspace{0.5cm}
\begin{center}
{\Large \bf Quantum master equation for the vacuum decay dynamics}

\lineskip .75em
\vskip 2.5cm
{\large Hong Wang$^{a}$, Jin Wang$^{b,}$\footnote{jin.wang.1@stonybrook.edu} }
\vskip 2.5em
 {\normalsize\it $^{a}$State Key Laboratory of Electroanalytical Chemistry, Changchun Institute of Applied Chemistry, Chinese Academy of Sciences, Changchun 130022, China\\
 $^{b}$Department of Chemistry and Department of Physics and Astronomy, State University of New York at Stony Brook, NY 11794, USA}
\vskip 3.0em
\end{center}
\begin{abstract}
The quantum master equation required to describe the dynamics of gravity-related vacuum decay is still challenging. We aim to study this issue. Our model consists of the spacetime and scalar field with self-interaction potential. The environment is chosen as spacetime while the system is formed by the vacua of the scalar field. We demonstrate that the quantum dynamics of the vacua can be described by the Redfield equation, which can depict the evolution of both coherence and the comoving volume fraction of the vacuum. Under the Markovian limit, coherence monotonically decreases with time, leading to the initial quantum state to decohere into a classical state. This helps the understanding of the decoherence of the universe. We also highlight that in certain circumstances, the evolution of the vacuum system may display non-Markovian dynamics. In specific scenarios, the classical limit of the quantum master equation is consistent with the classical master equation. In the steady state, the dominant vacuum corresponds to the smallest cosmological constant, and various dS vacua can reach equilibrium states.
\end{abstract}
\end{titlepage}

\baselineskip=0.7cm

\tableofcontents
\newpage
\section{Introduction}
\label{sec:1}
The discovery of the gravitational waves by LIGO~\cite{BP}  have promoted the study to the subject of cosmological first-order phase transition. The stochastic gravitational wave background produced by this mechanism may be observed by experiments~\cite{ZA,SK,WZ,OJ}. This may aid in understanding the physics of the early universe.

Significant theoretical progresses have been made regarding the cosmological first-order phase transition induced by bubble nucleation. The essence of bubble nucleation is quantum tunneling. In 1980, Coleman and De Luccia studied the tunneling rate of vacuum decay related to gravity in the case where the temperature is zero~\cite{SF}. They have shown that the tunneling rate is determined by the Euclidean action of the bounce solution and the initial vacuum state. The bounce is composed of a pair of instanton-anti-instanton and has the O(4) symmetry~\cite{SF}. Subsequently, Linde considered vacuum decay for the case of non-zero temperature~\cite{AD1,AD2}. He showed that high temperatures promote vacuum tunneling. Both the Hartle-Hawking and the Vilenkin wave function of the universe indicate that the dS universe can tunnel from nothing~\cite{JH,AV1,AV2}. In 1990, several researchers  showed that the transition probability from the Minkowski vacuum to the dS vacuum is not equal to zero~\cite{EA,WD}.  Recently, Huang and Ford have pointed out that the vacuum radiation pressure  fluctuations may enhance the tunneling rate~\cite{HL1,HL2}. Other important works have also been undertaken regarding vacuum decay, see~\cite{SM,JC,SF2,JMH} and the references therein.

Conventionally, the dynamics of the vacuum system is often described by the classical master equation (or Pauli equation)~\cite{DS,ALI,MC,JL,JGA,HW,JG}
\begin{equation}
\label{eq:1.1}
\frac{dP_{i}(t)}{dt}=\sum_{j}\Gamma_{ij}P_{j}(t)-\sum_{j}\Gamma_{ji}P_{i}(t).
\end{equation}
Here, $P_{i}$ is the fraction of comoving volume of the vacuum corresponding to the cosmological constant $\Lambda_{i}$. And $\Gamma_{ij}$ is the tunneling rate from vacuum $\Lambda_{j}$ to vacuum $\Lambda_{i}$.  If the term $3H_{i}P_{i}(t)$ ($H_{i}$ is the Hubble constant of the universe $\Lambda_{i}$) is added to the right hand side of equation \eqref{eq:1.1}, then $P_{i}$ represents the proper volume fraction~\cite{ALI}. The evolution described by equation \eqref{eq:1.1} is typically irreversible. Additionally, as the universe dominated by the cosmological constant exponentially expands, the multiverse landscape may be emerged in the process of bubble nucleation~\cite{AH,AD3,LS}. Therefore, the classical master equation can be used to describe the evolution of the multiverse landscape~\cite{DS,JGA,ALI,LS}. Throughout this study, we use units where $16\pi G=\hbar=k_{B}=c=1$.

The vacuum system related to gravity may exhibit quantum natures under certain situations. For example, in the very early stages of the universe, the coherence, entanglement or other quantum natures of the vacuum system may have been significant. Consequently, the evolution of these quantities is important. It is evident that the classical master equation \eqref{eq:1.1} is inadequate for describing the evolution of these quantities or the transformation of the system from quantum to classical states. Currently, a quantum dynamic equation for the gravity related vacuum system is still under development.

In this work, we aim to develop the quantum master equation that can depict the quantum dynamics of the vacuum system.  Our model consists of the FRLW spacetime and a scalar field with a self-interaction potential, where the scalar field is minimally coupled to the spacetime. We have chosen the spacetime as the environment and the vacuum states of the scalar field as the system. Starting from the Wheeler-DeWitt equation, we argued that the von Neumann equation (trace out the environment) can be utilized to describe the partial dynamics of the quantum universe.  Strictly solving this equation is nearly impossible for the usual cases.  We have derived the quantum master equation based on the von Neumann equation, which can be employed to represent the quantum dynamics of the vacuum system.

The quantum master equation is valid for any number of vacuum states. For simplicity, we numerically simulated a system with two types of dS vacuum states.  We show that the coherence monotonically decrease over time under the Markovian limit, resulting in the transition from an initial quantum state to a final classical state.  This may contribute to our understanding of the decoherence of the universe. Additionally, we noted that the vacuum system's evolution can show non-Markovian behavior in certain cases. The absolute value of the flux between vacuum states also decreases over time, eventually approaching to zero. This indicates that in the steady state, the vacuum system is in an equilibrium state, and the detailed balance is preserved. In the steady state, we found that the dominant vacuum corresponds to the smallest cosmological constant.  We demonstrated that in certain cases, the classical limit of the quantum master equation is consistent with the classical master equation.


\section{Partial dynamics of the quantum universe}
\label{sec:2}
General relativity mandates that any isolated physical system must adhere to the generalized covariant principle. Taking the 3+1 decomposition for the spacetime, general covariance dictates that any isolated system must satisfy the subsequent constraints~\cite{CK}:
\begin{equation}
\label{eq:2.1}
H_{tot}=0;
\end{equation}
\begin{equation}
\label{eq:2.2}
H_{a}=0.
\end{equation}
Equations \eqref{eq:2.1} and \eqref{eq:2.2} are the Hamiltonian constraint and the diffeomorphism constraint, respectively. $H_{tot}$ represents the total Hamiltonian of the isolated system. These equations contain all the classical dynamic information of the system. The Dirac quantization procedure transform these constrains into~\cite{CK}
\begin{equation}
\label{eq:2.3}
\hat{H}_{tot}|\Psi\rangle=0;
\end{equation}
\begin{equation}
\label{eq:2.4}
\hat{H}_{a}|\Psi\rangle=0.
\end{equation}
Here, $|\Psi\rangle$ represents the wave function of the universe. Equation \eqref{eq:2.3} is the Wheeler-DeWitt equation. In the following, we have sometimes omitted the operator hat for convenience. Readers can easily distinguish between $c$-numbers and $q$-numbers based on the related content.

In equations \eqref{eq:2.3} and \eqref{eq:2.4}, there is no  time variable, making it difficult to extract quantum dynamical information about the entire universe. This is known as the time problem in quantum gravity, which has not yet been perfectly solved~\cite{CK,KV,CJ}.  For a homogenous and isotropic spacetime, the diffeomorphism constraints \eqref{eq:2.2} and \eqref{eq:2.4} are trivial and do not need to be considered. Equation \eqref{eq:2.3} can be written equivalently as~\cite{HW2}
\begin{equation}
\label{eq:2.5}
\frac{d\rho_{tot}}{dt}=-i[H_{tot}, \rho_{tot}]=0.
\end{equation}
Here, $\rho_{tot}$ is the density matrix of the universe.

The universe can be divided into two parts: the system and the environment. If we are just interested in the dynamical information of the system, we need to trace out the environment~\cite{HW2}. By using $\rho$ to represent the reduced density matrix of the system, then one can obtain the von Neumann equation~\cite{HC,HF}
\begin{equation}
\label{eq:2.6}
\frac{d\rho}{dt}=-i\mathrm{Tr}_{B}[H_{tot}, \rho_{tot}].
\end{equation}
Here, $\mathrm{Tr}_{B}$ represents the partial trace for the environment. The von Neumann equation \eqref{eq:2.6} forms the foundation of the  theory of open quantum systems. Although the general covariance requires that $d\rho_{tot}/dt=0$ for any isolated system, this does not imply that $d\rho/dt$ must also be equal to zero! Typically, as one part of the universe, the system interacts with the environment and can not be seen as an isolated one. Thus, the general covariance does not constrain the (effective) Hamiltonian of the system to be equal to zero~\cite{HW3}. Or in other words, the general covariance does not constrain $d\rho/dt$ to be equal to zero. Therefore, one can use equation \eqref{eq:2.6} to describe the quantum dynamics of the system~\cite{HW3}.

Looking at this from another perspective, the system interacts with the environment. It will evolve under the influence of the environment. Energy may flow between the system and the environment. Some of the physical quantities, such as the coherence and the entanglement entropy,  will change over time. This means that $d\rho/dt\neq0$, even for the systems related to gravity~\cite{HW3}. Extracting all the quantum dynamic information of the universe seems to be blocked by the general covariance according to the Wheeler-DeWitt equation \eqref{eq:2.3}, yet we can still obtain partial dynamical information of the quantum universe according to equation \eqref{eq:2.6}.

Starting from the Wheeler-DeWitt equation \eqref{eq:2.3}, one can also obtain the von-Neumann equation \eqref{eq:2.6} in another way. The action of the Brown-Kucha$\check{\mathrm{r}}$  dust field is~\cite{JD,VT,HW2}
\begin{equation}
\label{eq:2.7}
S_{\bm{T}}=-\frac{1}{2}\int dx^{4}\sqrt{-g}\rho (g^{\mu\nu}\partial_{\mu} \bm{T}\partial_{\nu} \bm{T}+1).
\end{equation}
Here,  $\rho$  and $\bm{T}$ represent the rest mass density and the dust field, respectively. $g^{\mu\nu}$ is the metric of the spacetime and $g$ is the determinant of the metric. Using the canonical time gauge fixing condition $\bm{T}=t$~\cite{VT,KC} (this gauge condition sets the dust field as the clock), one can obtain the Wheeler-DeWitt equation~\cite{VT}
\begin{equation}
\label{eq:2.8}
(P_{\bm{T}}+H_{tot})| \Psi \rangle=0.
\end{equation}
Here, $P_{\bm{T}}=-i\partial/\partial t$ is the Hamiltonian operator of the dust field and $H_{tot}$ is the total Hamiltonian of the universe (except for the dust field). Equation \eqref{eq:2.8} can be equivalently written as $d\rho_{tot}/dt=-i[H_{tot},\rho_{tot}]$. After tracing out the environment, one can obtain the von Neumann equation \eqref{eq:2.6}~\cite{HW2}. Therefore, one can also interpret the time variable in equation \eqref{eq:2.6} as the dust field.

Equation \eqref{eq:2.8} indicates that if the Hamiltonian of the dust field is non-zero, the clock will influence the evolution of the universe. This is unsatisfactory as it is generally expected in physical theory that the clock should have no impact on the system. One can use the constraint $\mathrm{Tr}(H_{tot}\rho_{tot})=0$ to eliminate the influence of the clock on the evolution of the universe. It is easily to prove that $\mathrm{Tr}(H_{tot}\rho_{tot})$ is conserved. Therefore, this constraint can typically be met by selecting the initial state of the universe appropriately~\cite{HW2}.

In 2015, Maeda used equation \eqref{eq:2.8} to study the quantum evolution of the dark energy dominated universe. He showed that the evolution of the wave packet is consistent with the classical trajectory of the universe~\cite{HM}. In~\cite{HW2}, based on equation \eqref{eq:2.6}, we also obtained the same conclusion for the radiation or the non-relativistic matter dominated universe. These works indicate that it is reasonable to study the quantum dynamics of the subsystem of the universe based on equation \eqref{eq:2.6}.

To sum up, equation \eqref{eq:2.6} can be derived from the Wheeler-DeWitt equation by introducing the Brown-Kucha$\check{\mathrm{r}}$ dust field. All of the above arguments indicate that it should be reasonable to use equation \eqref{eq:2.6} to describe the quantum evolution of the open gravitational system.  The more general form of this equation can be found in~\cite{HW3}. Next, based on equation \eqref{eq:2.6}, we will derive the quantum master equation for describing the vacuum decay quantum dynamics.

\section{The quantum master equation of the vacuum system}
\label{sec:3}
\subsection{The Hamiltonian of the model}
\label{sec:3.1}
\begin{figure}[tbp]
\centering
\includegraphics[width=8cm]{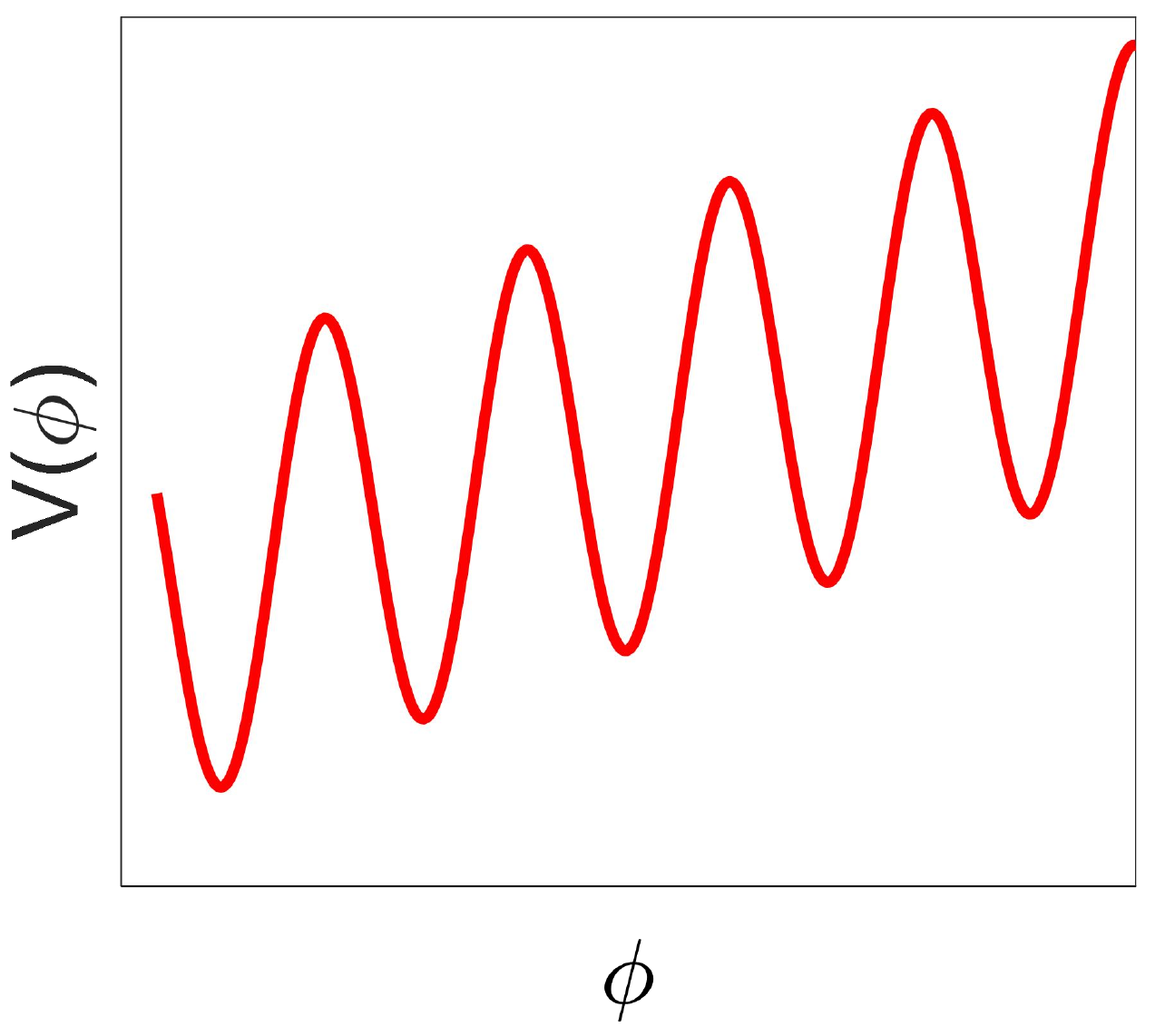}
\caption{\label{fig:1} The scalar potential diagram. The potential minimal state is the vacuum state.}
\end{figure}

The universe is composed of spacetime and matter fields. For simplicity, we assume that the matter is a scalar field. Then the total action of the universe is~\cite{SF}
\begin{eqnarray}\begin{split}
\label{eq:3.1}
S_{tot}&=S_{g}+S_{\phi}+S_{\partial \mathcal{M}}\\&=-\int dx^{4}\sqrt{-g}R +\int dx^{4}\sqrt{-g}(\frac{1}{2}g^{\mu\nu}\partial_{\mu}\phi\partial_{\nu}\phi-V(\phi))+S_{\partial \mathcal{M}}.
\end{split}
\end{eqnarray}
Here, $S_{tot}$ is the total action of the universe. $S_{g}$ and $S_{\phi}$ are the actions of the spacetime and scalar field, respectively. Matter and spacetime is for simplicity assumed to be minimally coupled. And $S_{\partial \mathcal{M}}$ is the Gibbons-Hawking surface term. $R$ is the Ricci scalar of the 4-dimensional spacetime and $\phi$ denotes the scalar field. $V(\phi)$ is the scalar potential. $V(\phi)$ has some minimal values. Or in other words, the scalar field has some non-degeneracy vacua, as shown in figure~\ref{fig:1}.

For the closed ($k=1$) homogenous and isotropic spacetime, the metric is~\cite{BM,VAR}
\begin{equation}
\label{eq:3.2}
ds^{2}=-dt^{2}+a^2(t)\big(d\chi^{2}+sin^{2}\chi (d\theta^{2}+sin^{2}\theta d\varphi^{2})\big).
\end{equation}
And the Ricci scalar is~\cite{BM}
\begin{equation}
\label{eq:3.3}
R=6\big(\frac{\ddot{a}}{a}+(\frac{\dot{a}}{a})^{2}+\frac{1}{a^{2}}\big),
\end{equation}
where $\dot{a}=da/dt$. Combining equations \eqref{eq:3.1}, \eqref{eq:3.2} and \eqref{eq:3.3}, the action of the spacetime becomes
\begin{equation}
\label{eq:3.4}
S_{g}=-\int dx^{4}\sqrt{-g}R=-12\pi^2\int dt(a\dot{a}^{2}-a).
\end{equation}
In equation \eqref{eq:3.4}, the boundary term generated by the partial integral for the coordinate time variable is eliminated by the Gibbons-Hawking surface term when going from the first to the second step.  Combining equations \eqref{eq:3.1} and \eqref{eq:3.2}, the action of the scalar field becomes
\begin{equation}
\label{eq:3.5}
S_{\phi}=\int 2\pi^{2} a^{3}\big(\frac{1}{2}\dot{\phi}^{2}-V(\phi)\big)dt.
\end{equation}
Noted that $2\pi^{2}a^{3}$ is the area of the 3-dimensional spherical surface which the sphere radius is $a$.  From equation \eqref{eq:3.4} and \eqref{eq:3.5}, one can obtain the Lagrangian of the spacetime and the scalar field are
\begin{equation}
\label{eq:3.6}
L_{g}=-12\pi^{2}(a\dot{a}^{2}-a)
\end{equation}
and
\begin{equation}
\label{eq:3.7}
L_{\phi}=2\pi^{2}a^{3}\big(\frac{1}{2}\dot{\phi}^{2}-V(\phi)\big),
\end{equation}
respectively.

From equations \eqref{eq:3.6} and \eqref{eq:3.7}, one can obtain the conjugate momentum of the scale factor $a$ and the scalar field $\phi$ as $P_{a}=\partial L_{g}/\partial \dot{a}=-24\pi^{2}a\dot{a}$ and $P_{\phi}=\partial L_{\phi}/\partial \dot{\phi}=2\pi^{2}a^{3}\dot{\phi}$, respectively. Then one can obtain the Hamiltonian of the spacetime
\begin{equation}
\label{eq:3.8}
H_{g}=P_{a}\dot{a}-L_{g}=-12\pi^{2}(a\dot{a}^{2}+a).
\end{equation}
The term $12\pi^{2}a$ in \eqref{eq:3.8} leads  the  classical dynamics of the closed ($k=1$) universe to be different from the flat ($k=0$) universe.  And the Hamiltonian of the scalar field in the FRLW spacetime is
\begin{equation}
\label{eq:3.9}
H_{\phi}=P_{\phi}\dot{\phi}-L_{\phi}=\frac{P_{\phi}^{2}}{4\pi^{2}a^{3}}+2\pi^{2}a^{3}V(\phi).
\end{equation}
Equation \eqref{eq:3.9} shows that every term in $H_{\phi}$ is coupled with the scale factor $a$. Or in other words, the free Hamiltonian of the scalar field  is zero. The coupling between the scalar field and the spacetime is non-linear. This is caused by the minimal coupling of the scalar field with the spacetime.  $H_{\phi}$ represents the interaction between the scalar field and  spacetime and is therefore sometimes referred to as the interaction Hamiltonian.

The total Hamiltonian of the universe is
\begin{eqnarray}\begin{split}
\label{eq:3.10}
H_{tot}&=H_{g}+H_{\phi}\\&=-12\pi^{2}(a\dot{a}^{2}+a)+\frac{P_{\phi}^{2}}{4\pi^{2}a^{3}}+2\pi^{2}a^{3}V(\phi).
\end{split}
\end{eqnarray}
The general covariance constrains $H_{tot}=H_{g}+H_{\phi}=0$ . This is the Friedmann equation. Solving this equation gives the classical dynamical information of the universe. However, general covariance does not restrict $H_{\phi}$ from being zero. Thus we expected that equation \eqref{eq:2.6} can be used to describe the quantum evolution of the scalar field. In addition, equation \eqref{eq:3.10} shows that $H_{\phi}$ cannot be considered a small quantity when compared to $H_{g}$.

We are interested in the quantum dynamics of the scalar field, so we have chosen the scalar field as the system and the spacetime as the environment. Carrying out the canonical quantization procedure to the universe, $P_{a}$ and $P_{\phi}$ are replaced by $\hat{P}_{a}=-i\partial /\partial a$ and $\hat{P}_{\phi}=-i\partial/\partial \phi$, respectively. The quantum dynamics of the scalar field is then determined by the von Neumann equation \eqref{eq:2.6}. After tracing out the spacetime, one can obtain the quantum master equation of the scalar field. This is the task of the next subsection.

\subsection{The quantum master equation}
\label{sec:3.2}
Equation \eqref{eq:2.6} determined the evolution of the system. Given the initial reduced density matrix, one can obtain the quantum dynamical information of the system in any time. Thus one can introduce the dynamical map $ \mathscr{V}(t,t_{0})$~\cite{HF} so that $\rho(t)=\mathscr{V}(t,t_{0})\rho(t_{0})$ where $\rho(t)$ and $\rho(t_{0})$ are the reduced density matrix of the system at the moment $t$ and $t_{0}$ ($t>t_{0}$), respectively.
Then, the time derivative of the reduced density matrix can be expressed as
\begin{equation}
\label{eq:3.11}
\frac{d\rho}{dt}=\lim_{\varepsilon\rightarrow 0}\frac{1}{\varepsilon}\big(\mathscr{V}(t+\varepsilon,t)\rho-\rho\big).
\end{equation}
The dynamical map $\mathscr{V}(t,t_{0})$ is a linear super-operator.

Assuming that there is no entanglement between the system and the environment in the initial state, and the set of orthonormal operators $\{\mathbf{F}_{i}, i=1,2,3,...; \mathrm{Tr}(\mathbf{F}_{i}^{\dag}\mathbf{F}_{j})=\delta_{ij}\}$ constitutes a complete basis of the Liouville space. Then $\mathscr{V}(t,t_{0})\rho$ can be expressed as~\cite{HF}
\begin{equation}
\label{eq:3.12}
\mathscr{V}(t,t_{0})\rho(t_{0})=\sum_{i,j=1}^{N^{2}}c_{ij}(t,t_{0})\mathbf{F}_{i}\rho(t_{0}) \mathbf{F}_{j}^{\dag}
\end{equation}
in the Liouville space. Here, $c_{ij}$ are $c$-number coefficients.   And $N$ is the dimension of the Hilbert space of the system. Bringing equation \eqref{eq:3.12} into \eqref{eq:3.11}, one can obtain the general form of the quantum master equation~\cite{HF}
\begin{equation}
\label{eq:3.13}
\frac{d\rho}{dt}=-i[H_{FL}, \rho] +D(\rho)
\end{equation}
where
\begin{equation}
\label{eq:3.14}
H_{FL}=\frac{1}{2\sqrt{N}i}\sum_{j=1}^{N^{2}-1}\big(\lim_{\varepsilon\rightarrow 0}\frac{c_{jN^{2}}^{*}(t+\varepsilon, t)}{\varepsilon}\mathbf{F}_{j}^{\dag}-\lim_{\varepsilon\rightarrow 0}\frac{c_{jN^{2}}(t+\varepsilon, t)}{\varepsilon}\mathbf{F}_{j}\big),
\end{equation}
\begin{equation}
\label{eq:3.15}
D(\rho)=\sum_{i,j=1}^{N^{2}-1}\lim_{\varepsilon\rightarrow 0}\frac{c_{ij}(t+\varepsilon, t)}{\varepsilon}\big(\mathbf{F}_{i}\rho \mathbf{F}_{j}^{\dag}-\frac{1}{2}\{\mathbf{F}_{j}^{\dag}\mathbf{F}_{i}, \rho\}\big).
\end{equation}
Here, $\{A,B\}\equiv AB+BA$. In equation \eqref{eq:3.14}, $H_{FL}$ is the sum of the free Hamiltonian and the Lamb shift Hamiltonian of the system~\cite{HF,FM}. In equation \eqref{eq:3.15}, $D(\rho)$ represents the dissipator. $D(\rho)$ leads to non-unitary quantum evolution in the system.

In the model where the total Hamiltonian is defined by equation \eqref{eq:3.10}, the free Hamiltonian of the system (scalar field) is zero. If we neglect the Lamb shift Hamiltonian, we will have $H_{FL}=0$. Moreover, dissipation, irreversible non-equilibrium evolution, decoherence and other non-unitary related behaviors are all described by the dissipator $D(\rho)$. These are induced by the interaction between the system and the environment. If we define  $A=2\pi^{2}a^{3}$ and $K=P_{\phi}^{2}/2$, the interaction Hamiltonian operator $H_{\phi}$ becomes
\begin{equation}
\label{eq:3.16}
H_{\phi}=A^{-1}K+AV.
\end{equation}
Here, $K$ can be seen as the kinetic energy operator of the scalar field and $V$ is the potential energy operator. Equation \eqref{eq:3.16} indicates that all the non-unitary behaviors in the system are induced by the operators $K$  and $V$ ($A^{-1}$ and $A$ are operators of the environment. After tracing out the environment, they are absorbed into the $c$-number coefficients of the dissipator ) . Thus, it is reasonable to infer that the dissipator is constructed using the operators $K$ and $V$, that is
\begin{eqnarray}\begin{split}
\label{eq:3.17}
D(\rho)=&a_{11}\big(K\rho K^{\dag}-\frac{1}{2}\{K^{\dag}K,\rho\}\big)+ a_{12}\big(K\rho V^{\dag}-\frac{1}{2}\{V^{\dag}K,\rho\}\big)\\&+ a_{21}\big(V\rho K^{\dag}-\frac{1}{2}\{K^{\dag}V,\rho\}\big)+ a_{22}\big(V\rho V^{\dag}-\frac{1}{2}\{V^{\dag}V,\rho\}\big).
\end{split}
\end{eqnarray}
Here, $a_{ij}$ are $c$-number coefficients. To Summarize these arguments, the form of the quantum master equation should be
\begin{eqnarray}\begin{split}
\label{eq:3.18}
\frac{d\rho}{dt}=&a_{11}\big(K\rho K^{\dag}-\frac{1}{2}\{K^{\dag}K,\rho\}\big)+ a_{12}\big(K\rho V^{\dag}-\frac{1}{2}\{V^{\dag}K,\rho\}\big)\\&+ a_{21}\big(V\rho K^{\dag}-\frac{1}{2}\{K^{\dag}V,\rho\}\big)+ a_{22}\big(V\rho V^{\dag}-\frac{1}{2}\{V^{\dag}V,\rho\}\big).
\end{split}
\end{eqnarray}

It is easy to prove that the operators $K$ and $V$ are Hermitian operators. However,  we still distinguish $K$ ($V$) and $K^{\dag}$ ($V^{\dag}$)  because we are interested in the coarse-grained dynamical information of the system, and coarse-graining may destroy the Hermiticity of the operator. The detailed contents about the coarse-graining will be presented in the next section. The quantum master equation \eqref{eq:3.18} determines the quantum dynamics of the scalar field. It is obtained above based on some arguments rather than rigorous derivation. Therefore,  before coarse-graining this quantum master equation, it is useful to derive it in a more rigorous way.

For convenience, we temporarily work in the interaction picture. In the interaction picture, the von Neumann equation \eqref{eq:2.6} becomes~\cite{HC,HF}
\begin{equation}
\label{eq:3.19}
\frac{d\widetilde{\rho}}{dt}=-i\mathrm{Tr}_{B}[\widetilde{H}_{\phi}, \widetilde{\rho}_{tot}],
\end{equation}
where
\begin{equation}
\label{eq:3.20}
\widetilde{\rho}=e^{iH_{g}t}\rho e^{-iH_{g}t},
\end{equation}
\begin{equation}
\label{eq:3.21}
\widetilde{H}_{\phi}=e^{iH_{g}t}H_{\phi} e^{-iH_{g}t},
\end{equation}
\begin{equation}
\label{eq:3.22}
\widetilde{\rho}_{tot}=e^{iH_{g}t}\rho_{tot} e^{-iH_{g}t}.
\end{equation}
$\widetilde{\rho}$ is the reduced density matrix of the system in the interaction picture. $\widetilde{H}_{\phi}$ is the interaction Hamiltonian in the interaction picture. And $\widetilde{\rho}_{tot}$ is the density matrix of the universe in the interaction picture. It should be noted that as the free Hamiltonian of the scalar field is zero, the free Hamiltonian of the total system is $H_{g}$. For the same reason, $\widetilde{K}=K$ and $\widetilde{V}=V$ where $\widetilde{K}$ and $\widetilde{V}$ are the kinetic and the potential energy operators of the scalar field in the interaction picture, respectively. In addition, $\widetilde{A}=e^{iH_{g}t}A e^{-iH_{g}t}$ and $\widetilde{A}^{-1}=e^{iH_{g}t}A^{-1} e^{-iH_{g}t}$. Thus, one can obtain
\begin{equation}
\label{eq:3.23} 
\widetilde{H}_{\phi}=\widetilde{A}^{-1}\widetilde{K}+\widetilde{A}\widetilde{V}.
\end{equation}

Equation \eqref{eq:3.19} can be equivalently written as~\cite{HC,HF}
\begin{equation}
\label{eq:3.24}
\frac{d\widetilde{\rho}(t)}{dt}=-i\mathrm{Tr}_{B}[\widetilde{H}_{\phi}(t), \widetilde{\rho}_{tot}(0)]-\int_{0}^{t}ds \mathrm{Tr}_{B} [\widetilde{H}_{\phi}(t),[\widetilde{H}_{\phi}(s),\widetilde{\rho}_{tot}(s)]],
\end{equation}
where $\widetilde{\rho}_{tot}(0)$ is the initial state of the universe in the interaction picture. It is common to set $\mathrm{Tr}_{B}(\widetilde{H}_{\phi}(t) \widetilde{\rho}_{tot}(0))=0$ when studying the dynamics of the open quantum system~\cite{HC,HF}. Noted that the classical master equation is a Markovian master equation. That is, $dP_{i}(t)/dt$ only depends on $P_{i}(t)$ and not on $P_{i}(s) (s\neq t)$. In addition, $P_{i}(t)$ corresponds to the diagonal element of $\widetilde{\rho}(t)$. Thus, it seems reasonable to introduce the Markov approximation (replacing $\widetilde{\rho}_{tot}(s)$ with $\widetilde{\rho}_{tot}(t)$~\cite{HF}) in equation \eqref{eq:3.24}. Then equation \eqref{eq:3.24} becomes  the Redfield equation~\cite{HC,HF}
\begin{equation}
\label{eq:3.25}
\frac{d\widetilde{\rho}(t)}{dt}=-\int_{0}^{t}ds \mathrm{Tr}_{B} [\widetilde{H}_{\phi}(t),[\widetilde{H}_{\phi}(s),\widetilde{\rho}_{tot}(t)]].
\end{equation}
From equation \eqref{eq:3.24} to \eqref{eq:3.25}, although the Markov approximation is introduced, this does not mean that \eqref{eq:3.25} represents a Markovian quantum master equation~\cite{HF}.

In this work, we consider a specific case where the entanglement between the system and the environment is small and can be neglected. Then the density matrix of the universe can be written as
\begin{equation}
\label{eq:3.26}
\widetilde{\rho}_{tot}(t)\approx \widetilde{\rho}(t)\otimes\widetilde{\rho}_{B}(t).
\end{equation}
Here, $\widetilde{\rho}_{B}(t)$ represents the density matrix of the environment in the interaction picture. We pointed out that equation \eqref{eq:3.26} differs from the usual Born approximation. In the Born approximation, $\widetilde{\rho}_{B}$ remains unchanged with time~\cite{HC,HF}. However, in equation \eqref{eq:3.26}, $\widetilde{\rho}_{B}(t)$ may change over time. The Born approximation is only valid when the interaction Hamiltonian is enough small compared to the free Hamiltonian. Equation \eqref{eq:3.10} clearly shows that the interaction Hamiltonian $H_{\phi}$ can not be seen as a small quantity. Thus the Born approximation cannot be applied.

Bringing equation \eqref{eq:3.26} into \eqref{eq:3.25} and using the property  $\widetilde{H}_{\phi}=\widetilde{H}_{\phi}^{\dag}$, then equation \eqref{eq:3.25} becomes
\begin{eqnarray}\begin{split}
\label{eq:3.27}
\frac{d\widetilde{\rho}}{dt}=-\int_{0}^{t}ds & \mathrm{Tr}_{B}\Big\{\widetilde{H}_{\phi}^{\dag}(t)\widetilde{H}_{\phi}(s)\widetilde{\rho}(t)\otimes\widetilde{\rho}_{B}(t)-\widetilde{H}_{\phi}(t)\widetilde{\rho}(t)\otimes\widetilde{\rho}_{B}(t)\widetilde{H}_{\phi}^{\dag}(s)\\&-\widetilde{H}_{\phi}(s)\widetilde{\rho}(t)\otimes\widetilde{\rho}_{B}(t)\widetilde{H}_{\phi}^{\dag}(t)+\widetilde{\rho}(t)\otimes\widetilde{\rho}_{B}(t)\widetilde{H}_{\phi}^{\dag}(s)\widetilde{H}_{\phi}(t)\Big\}.
\end{split}
\end{eqnarray}
Combining equations \eqref{eq:3.23} and \eqref{eq:3.27}, one can obtain
\begin{eqnarray}\begin{split}
\label{eq:3.28}
\frac{d\widetilde{\rho}}{dt}=&2\int_{0}^{t}ds \mathrm{Tr}_{B}\Big(\widetilde{A}^{-1}(t)\widetilde{A}^{-1}(s)\widetilde{\rho}_{B}(t)\Big)\Big(\widetilde{K}\widetilde{\rho} \widetilde{K}^{\dag}-\frac{1}{2}\{\widetilde{K}^{\dag}\widetilde{K},\widetilde{\rho}\}\Big)\\&+ 2\int_{0}^{t}ds \mathrm{Tr}_{B}\Big(\widetilde{A}^{-1}(t)\widetilde{A}(s)\widetilde{\rho}_{B}(t)\Big)\Big(\widetilde{K}\widetilde{\rho} \widetilde{V}^{\dag}-\frac{1}{2}\{\widetilde{V}^{\dag}\widetilde{K},\widetilde{\rho}\}\Big)\\&+ 2\int_{0}^{t}ds \mathrm{Tr}_{B}\Big(\widetilde{A}^{-1}(t)\widetilde{A}(s)\widetilde{\rho}_{B}(t)\Big)\Big(\widetilde{V}\widetilde{\rho} \widetilde{K}^{\dag}-\frac{1}{2}\{\widetilde{K}^{\dag}\widetilde{V},\widetilde{\rho}\}\Big)\\&+ 2\int_{0}^{t}ds \mathrm{Tr}_{B}\Big(\widetilde{A}(t)\widetilde{A}(s)\widetilde{\rho}_{B}(t)\Big)\Big(\widetilde{V}\widetilde{\rho} \widetilde{V}^{\dag}-\frac{1}{2}\{\widetilde{V}^{\dag}\widetilde{V},\widetilde{\rho}\}\Big).
\end{split}
\end{eqnarray}
We introduce the following definitions:
\begin{equation}
\label{eq:3.29}
a_{11}\equiv 2\int_{0}^{t}ds \mathrm{Tr}_{B}\Big(\widetilde{A}^{-1}(t)\widetilde{A}^{-1}(s)\widetilde{\rho}_{B}(t)\Big),
\end{equation}
\begin{equation}
\label{eq:3.30}
a_{12}\equiv  2\int_{0}^{t}ds \mathrm{Tr}_{B}\Big(\widetilde{A}^{-1}(t)\widetilde{A}(s)\widetilde{\rho}_{B}(t)\Big),
\end{equation}
\begin{equation}
\label{eq:3.31}
a_{21}\equiv 2\int_{0}^{t}ds \mathrm{Tr}_{B}\Big(\widetilde{A}^{-1}(t)\widetilde{A}(s)\widetilde{\rho}_{B}(t)\Big),
\end{equation}
\begin{equation}
\label{eq:3.32}
a_{22}\equiv 2\int_{0}^{t}ds \mathrm{Tr}_{B}\Big(\widetilde{A}(t)\widetilde{A}(s)\widetilde{\rho}_{B}(t)\Big).
\end{equation}
It is obvious that $a_{ij}$ depends on the state of the environment. Thus different initial states of the environment correspond to different $a_{ij}$. Equations \eqref{eq:3.30} and \eqref{eq:3.31} show that $a_{12}=a_{21}$.  Using these definitions, equation \eqref{eq:3.28} can be written as
\begin{eqnarray}\begin{split}
\label{eq:3.33}
\frac{d\widetilde{\rho}}{dt}=&a_{11}\Big(\widetilde{K}\widetilde{\rho} \widetilde{K}^{\dag}-\frac{1}{2}\{\widetilde{K}^{\dag}\widetilde{K},\widetilde{\rho}\}\Big)+ a_{12}\Big(\widetilde{K}\widetilde{\rho} \widetilde{V}^{\dag}-\frac{1}{2}\{\widetilde{V}^{\dag}\widetilde{K},\widetilde{\rho}\}\Big)\\&+ a_{21}\Big(\widetilde{V}\widetilde{\rho} \widetilde{K}^{\dag}-\frac{1}{2}\{\widetilde{K}^{\dag}\widetilde{V},\widetilde{\rho}\}\Big)+ a_{22}\Big(\widetilde{V}\widetilde{\rho} \widetilde{V}^{\dag}-\frac{1}{2}\{\widetilde{V}^{\dag}\widetilde{V},\widetilde{\rho}\}\Big).
\end{split}
\end{eqnarray}

Equation \eqref{eq:3.33} is the quantum master equation in the interaction picture. Noted that  $\widetilde{K}=K$ and $\widetilde{V}=V$. In addition, the free Hamiltonian of the scalar field is zero. Thus, transforming equation \eqref{eq:3.33} into the Schr\"{o}dinger picture, it becomes equation \eqref{eq:3.18}. Equations \eqref{eq:3.29}, \eqref{eq:3.30}, \eqref{eq:3.31} and \eqref{eq:3.32} show that $a_{ij}$ may change with time. Thus, equation \eqref{eq:3.18} may be a non-Markovian quantum master equation~\cite{ID,HE}. The dynamics of the system may exhibit certain non-Markovian properties. In the case where $a_{ij}$ is unchanged with time, equation \eqref{eq:3.18} is a Markovian quantum master equation and the related dynamical map $\mathscr{V}$ forms the dynamical semigroup. Because we have not introduced the secular approximation, equation \eqref{eq:3.18} can also be referred to as the Redfield equation. The quantum dynamics of the scalar field is described by the Redfield equation \eqref{eq:3.18}.

Equation \eqref{eq:3.18} determines the quantum evolution of the scalar field, but is difficult to solve. In this work, we focus on a specific situation in which the probability of the scalar field being in the vacuum states is high. Thus all the other non-vacuum states are not important and can be neglected. This condition can be easily achieved by controlling the parameters in the potential $V(\phi)$.  In this case, the dynamics of the scalar field is reduced to the dynamics of the vacua. And the physical Hilbert space of the system is spanned by the vacuum states.

Equation \eqref{eq:3.18} not only can describe the evolution of the comoving volume fraction of the vacuum but also can be used to study the evolution of specific quantum properties of the system, such as coherence.  Equation \eqref{eq:3.18} can also describe the evolution of the quantum superposition state. However, the classical master equation \eqref{eq:1.1} can only describe the evolution of the comoving volume fraction.

To summarize, in this subsection, we derived the quantum master equation \eqref{eq:3.18} using two different methods. In the first method, in order to obtain \eqref{eq:3.18}, we neglected the Lamb shift Hamiltonian without the justification. However, in the second method, we did not make this approximation and still obtained the same equation. This suggests that neglecting the Lamb shift Hamiltonian is a conservative approach.

\subsection{Coarse graining}
\label{sec:3.3}
\begin{figure}[tbp]
\centering
\includegraphics[width=8cm]{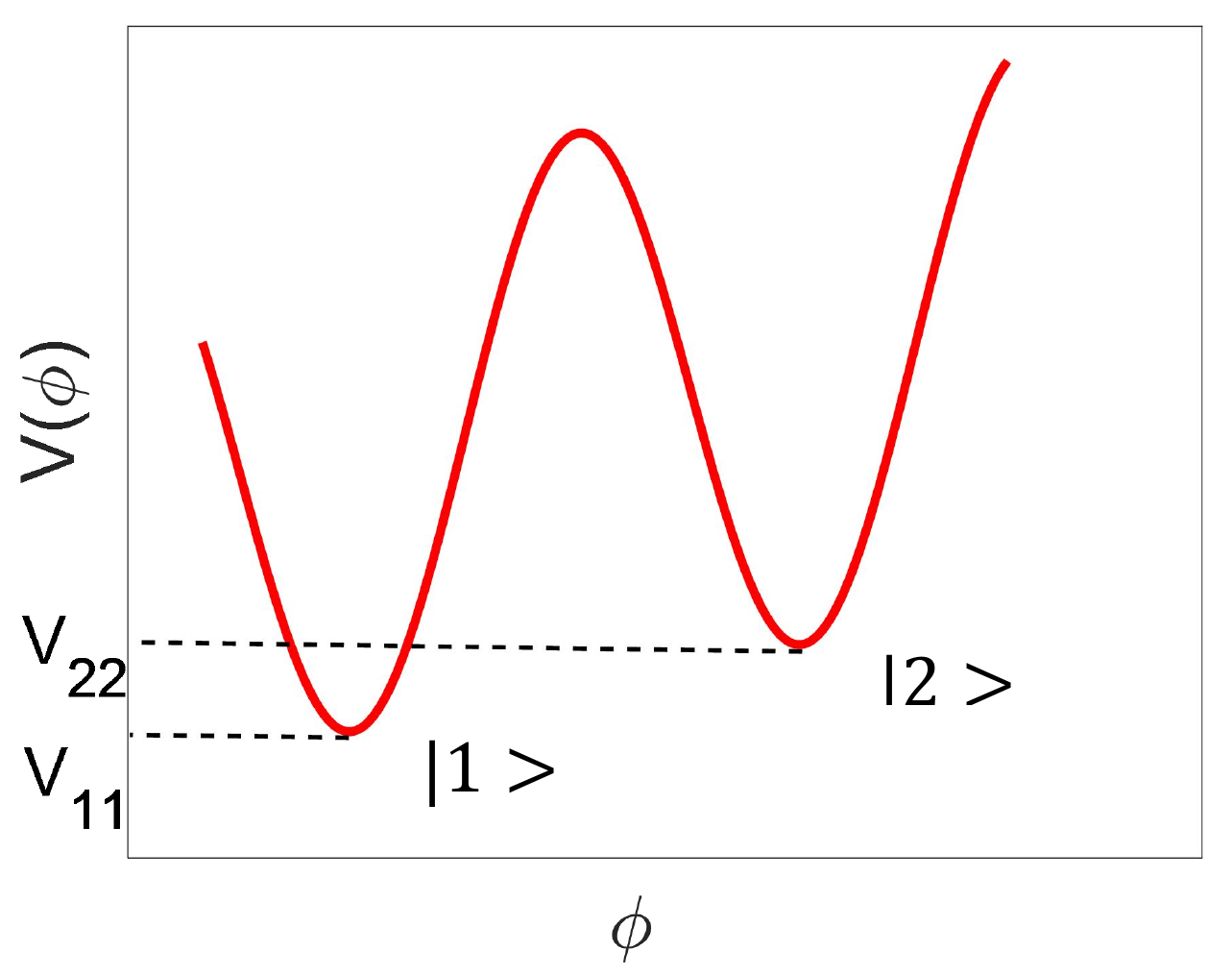}
\caption{\label{fig:2} The double well scalar potential. $V_{11}$ is the first minimal value and $V_{22}$ is the second minimal value. $|1\rangle$ is the first coarse grained vacuum state. $|2\rangle$ is the second coarse-grained vacuum state.}
\end{figure}

For simplicity, we consider a special case where there are only two potential minimal states, both of which are greater than zero, as shown in figure~\ref{fig:2}. Thus, the potential minimal states  correspond to the dS spaceime. It is well known that the temperature of the dS spacetime is~\cite{AL}
\begin{equation}
\label{eq:3.34}
T=\frac{1}{2\pi}\sqrt{\frac{\Lambda}{6}}
\end{equation}
where $\Lambda$ is the cosmological constant. $\Lambda$ is equal to the minimal value of the potential $V(\phi)$. Equation \eqref{eq:3.34} relates to the unit $16\pi G=1$. If we set $8\pi G=1$, then $T=\sqrt{\Lambda/12\pi^{2}}$. In addition, the Euclidean action of the dS spacetime is~\cite{DS,HW}
\begin{equation}
\label{eq:3.35}
S_{E}(\Lambda)=-\frac{16\pi^{2}}{H^{2}}=-\frac{96\pi^{2}}{\Lambda}.
\end{equation}
Here, $H=\sqrt{\Lambda/6}$ is the Hubble constant (If we set $8\pi G=1$, then $H=\sqrt{\Lambda/3}$ and $S_{E}(\Lambda)=-8\pi^{2}/H^{2}$). Combining equations \eqref{eq:3.34} and \eqref{eq:3.35}, one can obtain that the thermal entropy of the dS spacetime  is
\begin{eqnarray}\begin{split}
\label{eq:3.36}
\mathbf{S}(\Lambda)=\frac{U}{T}+\mathrm{ln}Z=\frac{1}{T}\cdot\frac{4}{3}\pi H^{-3}\cdot\Lambda-S_{E}(\Lambda)=-2S_{E}(\Lambda).
\end{split}
\end{eqnarray}
Here, $U$ and $Z$ are the internal energy and the partition function of the dS spacetime, respectively.

The relationship between the thermal entropy and the number of microstates is $\mathbf{S}=\mathrm{ln}\Omega$, where $\Omega$ represents the number of microstates. Thus, equation \eqref{eq:3.36} shows that the number of microstates of the dS spacetime is
\begin{equation}
\label{eq:3.37}
\Omega=e^{-2S_{E}(\Lambda)}.
\end{equation}
All these microstates correspond to the same vacuum energy $\Lambda$. However, in this work, we are not interested in the differences between these microstates. Thus, we need to coarse-grain the physical Hilbert space. The microstates of the dS spacetime are coarse-grained into the same state. The coarse-grained physical Hilbert space is spanned by two coarse-grained vacuum states $\{|1\rangle,|2\rangle;\langle \mu|\nu\rangle=\delta_{\mu\nu}\}$. The state $|\mu\rangle$ ($\mu, \nu=1, 2$) represents the vacuum state in which the cosmological constant is $\Lambda_{\mu}$. And the degree of degeneracy of the state $|\mu\rangle$ is $\Omega=e^{-2S_{E}(\Lambda_{\mu})}$.

The degree of degeneracy is induced by coarse graining. We take a simple example to illustrate this point. Assuming that there are three states, the distribution of these states in the steady state is uniform. That is $(P_{1}, P_{2}, P_{3})=(1/3, 1/3, 1/3)$ where $P_{i}$ is the fraction of the $i$th state. If we do not distinguish between the second and third states, then there are only two states, and the steady state distribution becomes $(P_{1}, P_{2})=(1/3, 2/3)$. Noted that $P_{2}=2\times1/3$. The factor ``2" represents the degree of degeneracy induced by the coarse graining. Thus, coarse graining induces the degree of degeneracy for the states.

The quantum master equation \eqref{eq:3.18} also needs to be coarse-grained. More specifically, the operators $K$ and $V$ need to be coarse-grained. The coefficients $a_{ij}$ are determined by the environment. Coarse graining for the system does not change these coefficients.  In addition, $V=\sum_{\mu\nu}V_{\mu\nu}|\mu\rangle\langle \nu|$. As different vacuum states are orthogonal to each other, $\langle \mu|\nu\rangle=\delta_{\mu\nu}$, thus the coarse-grained potential operator is
\begin{equation}
\label{eq:3.38}
V=
\begin{pmatrix}
V_{11}&0\\
0&V_{22}
\end{pmatrix}
=
\begin{pmatrix}
\Lambda_{1}&0\\
0&\Lambda_{2}
\end{pmatrix}.
\end{equation}
$V_{11}$ ($V_{22}$) is the first (second) minimal value of the potential $V(\phi)$. As a comparison, the potential operator without coarse-graining is
\begin{equation}
\label{eq:3.39}
V=
\begin{pmatrix}
V_{11}& & & & &  \\
 &\ddots& & & &  \\
 & &V_{11}& & & \\
 & & &V_{22}& &  \\
 & & & &\ddots&  \\
 & & & & &V_{22}\\
\end{pmatrix}.
\end{equation}
Here, the number of $V_{11}$ ($V_{22}$) is $e^{-2S_{E}(V_{11})}$ ($e^{-2S_{E}(V_{22})}$). Thus, the dimension of operator \eqref{eq:3.39} is $e^{-2(S_{E}(V_{11})+S_{E}(V_{22}))}$. However, the dimension of the coarse-grained operator $V$ is 2.

If the dS spacetime has no microstate, then the system is equivalent to a two level system. Generally, for the two level system, the kinetic energy operator is approximately proportional to the Pauli matrix $\sigma_{x}$. And the proportional coefficient is $e^{-S_{cl}}$~\cite{AS}.  We use $S_{cl}$ to represent the Euclidean action of the instanton, which is half of the Euclidean action of the bounce (a pair of instanton--anti-instanton) solution~\cite{UH}. Thus, we have
\begin{equation}
\label{eq:3.40}
K\approx
\begin{pmatrix}
0&e^{-S_{cl}}\\
e^{-S_{cl}}&0
\end{pmatrix}.
\end{equation}
$K_{\mu\nu}$ represents the tunneling amplitude from the state $|\nu\rangle$ to the state $|\mu\rangle$~\cite{AS}. Equation \eqref{eq:3.40} shows that the kinetic energy operator does not induce tunneling within the same state. In actuality, tunneling within the same state contributes to the vacuum energy. However, this effect is typically minor when compared to the minimal values of the classical potential. Thus one often neglect this effect~\cite{AS}.

Considering that dS spacetime has microstates and the number of microstates are given by equation \eqref{eq:3.37}, then the dimension of the operator $K$ should be $e^{-2(S_{E}(V_{11})+S_{E}(V_{22}))}$. Thus, equation \eqref{eq:3.40} should be generalized to
\begin{equation}
\label{eq:3.41}
K\approx
\begin{pmatrix}
\mathbf{0}&\mathbf{\Xi}\\
\mathbf{\Xi}&\mathbf{0}
\end{pmatrix}.
\end{equation}
Here, $\mathbf{0}$ is the zero matrix with the dimension  $e^{-2S_{E}(V_{11})}$. And $\mathbf{\Xi}$ is defined as
\begin{equation}
\label{eq:3.42}
\mathbf{\Xi}\equiv
\begin{pmatrix}
e^{-S_{cl}}& \cdots &e^{-S_{cl}}\\
\vdots&\ddots&\vdots\\
e^{-S_{cl}}& \cdots &e^{-S_{cl}}
\end{pmatrix}.
\end{equation}
The dimension of the matrix $\mathbf{\Xi}$ is $e^{-2S_{E}(V_{22})}$. The physical significance of equation \eqref{eq:3.41} is that if the two states correspond to the same dS spacetime, then the tunneling amplitude between these two states is zero, otherwise, the tunneling amplitude is $e^{-S_{cl}}$.

In order to show how to coarse-grain the kinetic energy operator \eqref{eq:3.41}.  We first examine the coarse-graining of the classical master equation \eqref{eq:1.1}. We consider a simple example in which the system has $N'$ states and the transition matrix $\Gamma$ is
\begin{equation}
\label{eq:3.43}
\Gamma=
\begin{pmatrix}
\omega & \cdots &\omega\\
\vdots&\ddots&\vdots\\
\omega& \cdots &\omega
\end{pmatrix}.
\end{equation}
Noted that the diagonal elements of the transition matrix do not have an influence on the evolution of the system. Thus the diagonal elements are not important. The dimension of the transition matrix \eqref{eq:3.43} is $N'$. One can easily prove that the steady state distribution is $(P_{1}, P_{2},..., P_{N'})=(1/N', 1/N',...,1/N')$. If we do not distinguish the second to the $N'$th states (these states are seen as the same state), then the steady state distribution becomes $(P_{1}, P_{2})=(1/N',(N'-1)/N')$.

After coarse graining the second to the $N'$th states into one state, the degree of degeneracy of this state is $N'-1$. Then, the transition rate from the first state to this coarse-grained state is $(N'-1)\omega$. Thus, the transition matrix \eqref{eq:3.43} should be coarse-grained into
\begin{equation}
\label{eq:3.44}
\Gamma=
\begin{pmatrix}
\omega & \omega\\
(N'-1)\omega & \omega
\end{pmatrix}.
\end{equation}
In equation \eqref{eq:3.44}, $\Gamma_{21}=(N'-1)\omega $ where the factor $(N'-1)$ represents the degree of degeneracy. This factor is induced by the coarse graining of the system. The dimension of the transition matrix \eqref{eq:3.44} is 2. Bringing \eqref{eq:3.44} into the classical master equation \eqref{eq:1.1}, one can easily prove that the steady state distribution is $(P_{1}, P_{2})=(1/N',(N'-1)/N')$. This example shows that if the coarse graining of the system contributes a degree of degeneracy $D_{e}$ to a state, then the transition rate from other states to this coarse-grained state will enhance $D_{e}$ times. In addition, we point out that the transition matrix \eqref{eq:3.43} is Hermitian but \eqref{eq:3.44} is not, indicating that the coarse graining can destroy the Hermiticity.

The coarse graining of the kinetic energy operator $K$ is independent of the coefficients $a_{ij}$. In this case where $a_{12}=a_{21}=a_{22}=0$,  the quantum master equation \eqref{eq:3.18} becomes
\begin{equation}
\label{eq:3.45}
\frac{d\rho}{dt}=a_{11}\big(K\rho K^{\dag}-\frac{1}{2}\{K^{\dag}K,\rho\}\big).
\end{equation}
Then, the relationship between the transition rate $\Gamma_{\mu\nu}$ and the tunneling amplitude $K_{\mu\nu}$ is~\cite{HE}
\begin{equation}
\label{eq:3.46}
\Gamma_{\mu\nu}=a_{11}\langle \mu|K|\nu\rangle\langle \nu|K^{\dag}|\mu\rangle.
\end{equation}

Equation \eqref{eq:3.46} can be equivalently written as
\begin{equation}
\label{eq:3.47}
D_{e}\Gamma_{\mu\nu}=a_{11}\sqrt{D_{e}}\langle \mu|K|\nu\rangle\sqrt{D_{e}}\langle \nu|K^{\dag}|\mu\rangle.
\end{equation}
This equation indicates that if the transition rate $\Gamma_{\mu\nu}$ enhances $D_{e}$ times, then the tunneling amplitude $K_{\mu\nu}$ should be enhanced $\sqrt{D_{e}}$ times. Therefore, the kinetic energy operator \eqref{eq:3.41} should be coarse-grained into
\begin{equation}
\label{eq:3.48}
K=
\begin{pmatrix}
0&e^{-S_{cl}-S_{E}(V_{11})}\\
e^{-S_{cl}-S_{E}(V_{22})}&0
\end{pmatrix}.
\end{equation}
The dimension of the coarse-grained kinetic energy operator \eqref{eq:3.48} is 2. Noted that for $K_{12}=e^{-S_{cl}-S_{E}(V_{11})}$, the factor $e^{-S_{E}(V_{11})}$ is the square root of the degree of degeneracy of the vacuum state $|1\rangle$. And for $K_{21}=e^{-S_{cl}-S_{E}(V_{22})}$, the factor $e^{-S_{E}(V_{22})}$ is the square root of the degree of degeneracy of the vacuum state $|2\rangle$. Thus, the coarse-grained kinetic energy operator is not a Hermitian operator, $K\neq K^{\dag}$. This is the reason that we distinguish between $K$ and $K^{\dag}$ in the quantum master equation \eqref{eq:3.18}.

To sum up, in this section, we have coarse-grained the quantum master equation. If we consider the case where the potential has two minimal values, then the dimension of the coarse-grained physical Hilbert space is two. Thus, the dimensions of the coarse-grained reduced density matrix, kinetic energy operator and potential energy operator all are two. Coarse graining has destroyed the Hermiticity of the kinetic energy operator.  Next, we will simulate the dynamics of the vacuum system for certain special situations using the coarse-grained quantum master equation.

\section{Simulations for certain cases}
\label{sec:4}

The scalar potential differs among various models. For convenience, we consider the washboard scalar potential~\cite{FD}
\begin{equation}
\label{eq:4.1}
V(\phi)=\alpha \phi +\beta \mathrm{cos} \gamma\phi +\delta,
\end{equation}
where, $0<\phi<4\pi/\gamma$. $\alpha$, $\beta$ and $\gamma$ are some positive parameters. In this interval, the potential has two vacuum states. We constrain that $\delta>\beta$ and $\alpha/\gamma$ is small compared to $V(\phi)$. Under these constraints, the thin-wall approximation is valid and the energy of the vacuum states is greater than zero. Thus the potential minimal states are dS vacua. Equation \eqref{eq:4.1} shows that $\phi_{1}=\pi/\gamma$ corresponds to the first dS vacuum state $|1\rangle$, and $\phi_{2}=3\pi/\gamma$ corresponds to the second dS vacuum state $|2\rangle$. The cosmological constant of the first (second) dS vacuum is $\Lambda_{1}=V_{11}=\delta-\beta+\alpha\pi/\gamma$ ($\Lambda_{2}=V_{22}=\delta-\beta+3\alpha\pi/\gamma$). Noted that $\Lambda_{1}<\Lambda_{2}$. Thus, sometimes  $|1\rangle$   and $|2\rangle$ are  referred to as the true vacuum state and  false vacuum state, respectively.  As $\alpha/\gamma$ is a small quantity, then $V_{22}-V_{11}=2\alpha\pi/\gamma$ is also small. Under this condition, the thin-wall approximation is valid and the wall of the Euclidean bounce is small in thickness~\cite{SF}.

Introducing the definition
\begin{equation}
\label{eq:4.2}
U(\phi)\equiv\beta \mathrm{cos} \gamma\phi +\delta,
\end{equation}
then the potential \eqref{eq:4.1} can be written as~\cite{SF}
\begin{equation}
\label{eq:4.3}
V(\phi)=U(\phi)+o(\frac{\alpha}{\gamma}).
\end{equation}
Here, $o(\alpha/\gamma)$ is a small quantity. $U(\phi)$ is similar to the  axion field potential~\cite{BM}.  According to the definition \eqref{eq:4.2}, one can easily show that $U(\phi_{1})=U(\phi_{2})$. The dS vacuum is a bubble, and the surface tension of the bubble is~\cite{VAR}
\begin{eqnarray}\begin{split}
\label{eq:4.4}
\tau&=\int_{\phi_{1}}^{\phi_{2}}d\phi \sqrt{2\big(V(\phi)-V(\phi_{2})\big)}\\&\approx\int_{\phi_{1}}^{\phi_{2}}d\phi \sqrt{2\big(U(\phi)-U(\phi_{2})\big)}.
\end{split}
\end{eqnarray}
Bringing equation \eqref{eq:4.2} into \eqref{eq:4.4}, one can obtain
\begin{equation}
\label{eq:4.5}
\tau=2\gamma\sqrt{\beta}.
\end{equation}

In~\cite{SF2}, de Alwis and other researchers present the detailed calculation for the Euclidean action $S_{cl}$. The result is
\begin{equation}
\label{eq:4.6}
S_{cl}=-\pi^{2}\Big\{\frac{\big(16(H_{2}^{2}-H_{1}^{2})^{2}+\tau^{2}(H_{1}^{2}+H_{2}^{2})\big)R_{o}}{2\tau H_{1}^{2}H_{2}^{2}}-4(H_{1}^{-2}+H_{2}^{-2})\Big\},
\end{equation}
where
\begin{equation}
\label{eq:4.7}
R_{o}=\frac{64\tau^{2}}{16^{2}(H_{1}^{2}-H_{2}^{2})^{2}+32\tau^{2}(H_{1}^{2}+H_{2}^{2})+\tau^{4}}.
\end{equation}
In equations \eqref{eq:4.6} and \eqref{eq:4.7}, $H_{1}=\sqrt{V_{11}/6}$ and $H_{2}=\sqrt{V_{22}/6}$ are the Hubble constants of the dS vacuum  $|1\rangle$ and $|2\rangle$, respectively ( If we set $8\pi G=1$, then $H_{1}=\sqrt{V_{11}/3}$ and $H_{2}=\sqrt{V_{22}/3}$ ).

One can learn from equations \eqref{eq:3.29}-\eqref{eq:3.32} that the coefficients $a_{ij}$ are difficult to calculate. Thus, in this work, we consider some special cases. Even so, valuable insights into the quantum dynamics of the vacuum system can be gained. The classical master equation \eqref{eq:1.1} is a Markovian master equation. Hence, we investigate the Markovian limit of the quantum master equation \eqref{eq:3.18}. In this case, one can neglect the variation of the coefficients $a_{ij}$.  Bringing equations \eqref{eq:3.38}, \eqref{eq:3.48}, \eqref{eq:4.5}, \eqref{eq:4.6} and \eqref{eq:4.7} into \eqref{eq:3.18}, then one can simulate the evolution of the vacuum system.

\begin{figure}[tbp]
\centering
\subfigure[]{
\begin{minipage}[t]{0.48\textwidth}
\centering
\includegraphics[width=7cm]{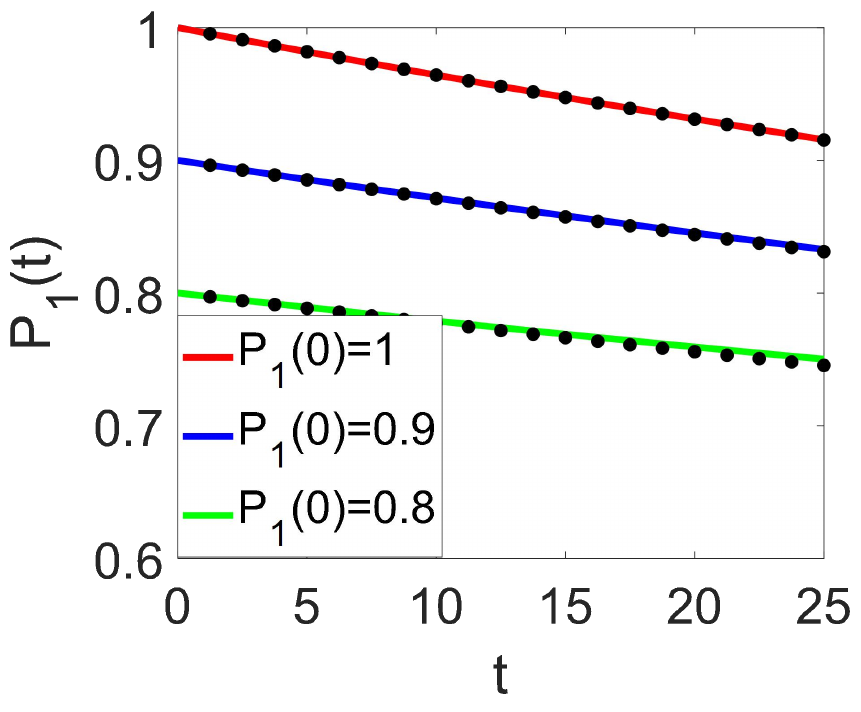}
\label{fig:3a}
\end{minipage}}
\subfigure[]{
\begin{minipage}[t]{0.48\textwidth}
\centering
\includegraphics[width=8cm]{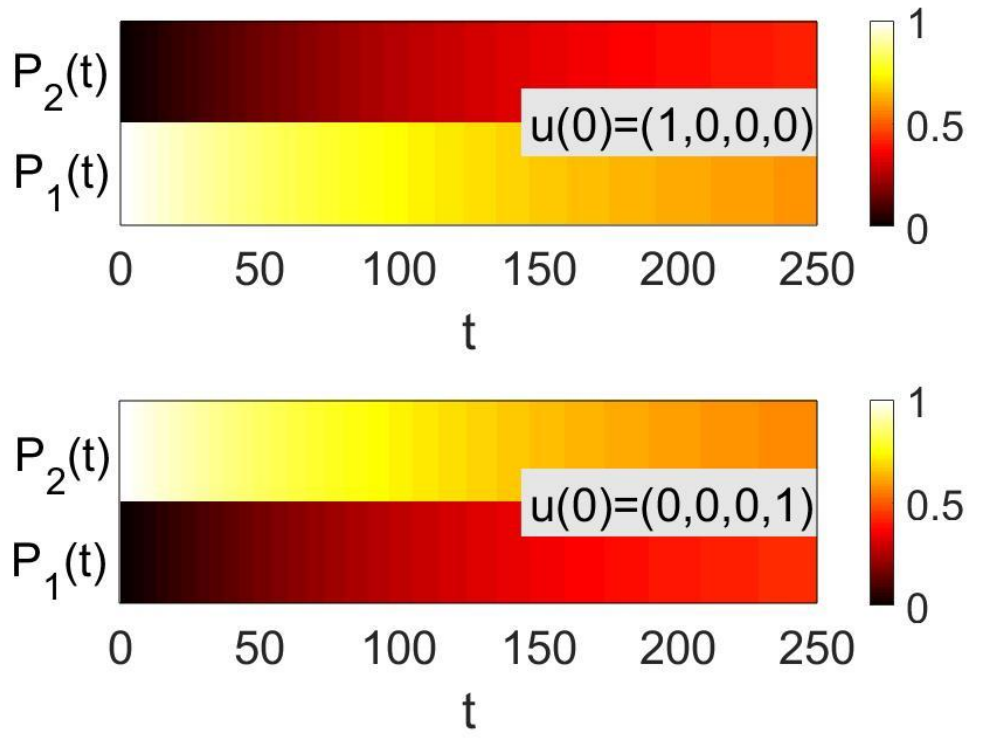}
\label{fig:3b}
\end{minipage}}
\subfigure[]{
\begin{minipage}[t]{0.48\textwidth}
\centering
\includegraphics[width=7cm]{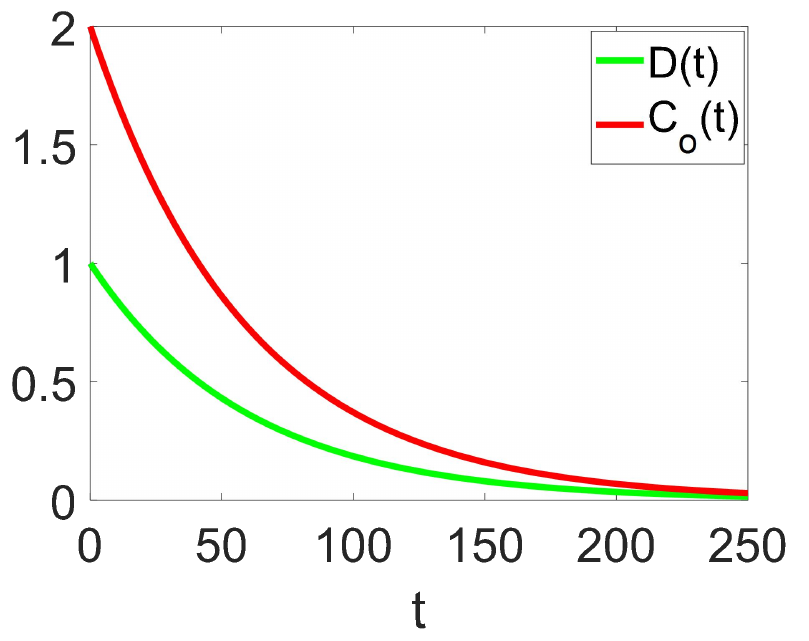}
\label{fig:3c}
\end{minipage}}
\subfigure[]{
\begin{minipage}[t]{0.48\textwidth}
\centering
\includegraphics[width=7cm]{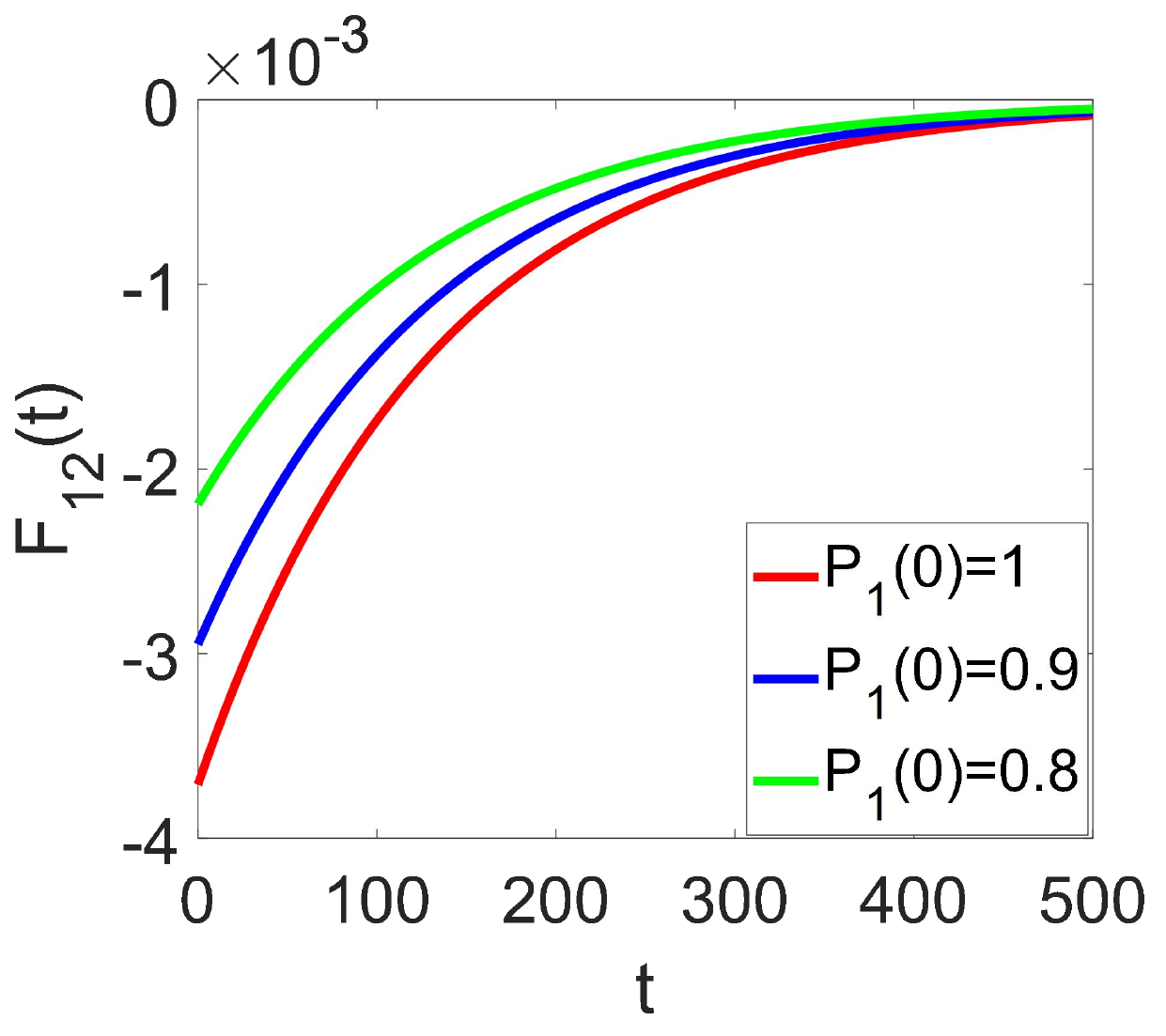}
\label{fig:3d}
\end{minipage}}
\caption{The evolution of certain quantities of the vacuum system. The horizontal axis of these figures are the time variable. In figure~\ref{fig:3a}, the vertical axis represents the fraction of the true vacuum. The solid lines correspond to the quantum master equation, and the dotted lines correspond to the classical master equation. In figure~\ref{fig:3b}, $P_{1}(t)$ and $P_{2}(t)$ represent the fraction of the true vacuum and the false vacuum, respectively. In figure~\ref{fig:3c}, $D(t)$ and $C_{o}(t)$ represent the decoherence function and the coherence, respectively. In figure~\ref{fig:3d}, $F_{12}(t)$ is the flux. The parameters are set as: $a_{11}=e^{2S_{E}(\Lambda_{1})+2S_{E}(\Lambda_{2})}$, $a_{12}=a_{21}=a_{22}=0$, $\alpha=1$, $\beta=1$, $\delta=500$. In figures~\ref{fig:3a}, ~\ref{fig:3b} and ~\ref{fig:3d}, $\gamma=1$. In figure~\ref{fig:3c}, $\gamma=0.02$. }
\label{fig:3}
\end{figure}

\begin{figure}[tbp]
\centering
\subfigure[]{
\begin{minipage}[t]{0.48\textwidth}
\centering
\includegraphics[width=7.8cm]{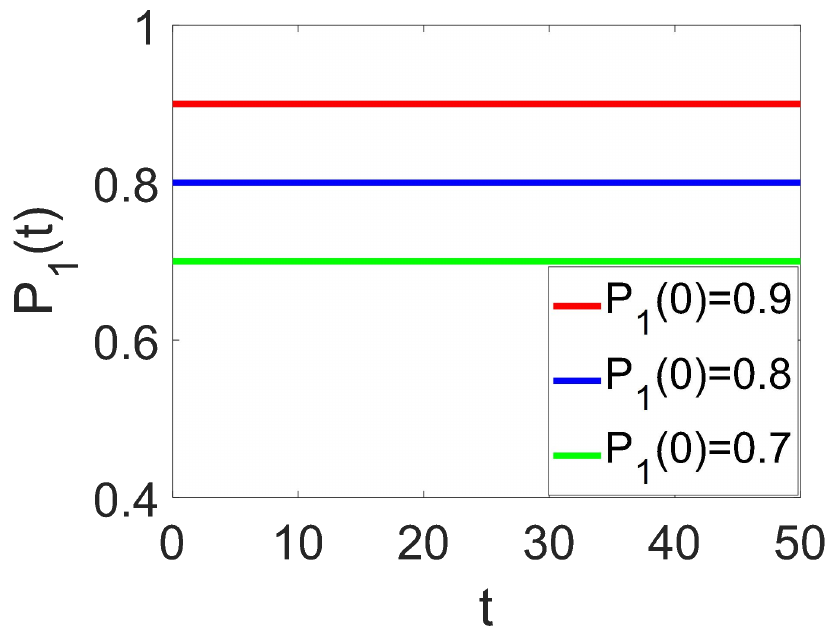}
\label{fig:4a}
\end{minipage}}
\subfigure[]{
\begin{minipage}[t]{0.48\textwidth}
\centering
\includegraphics[width=7cm]{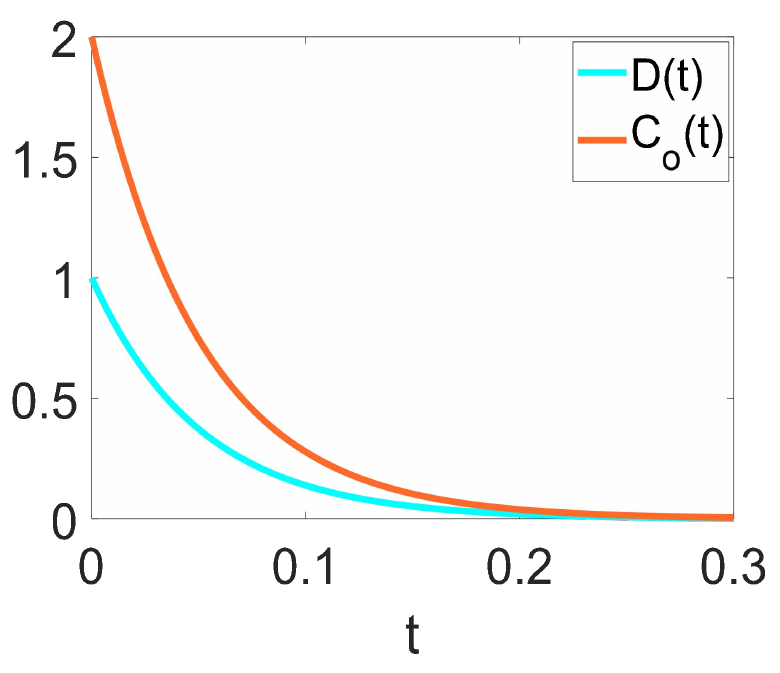}
\label{fig:4b}
\end{minipage}}
\caption{The evolution of certain quantities of the vacuum system. The horizontal axis of these two figures are the time variable. In figure~\ref{fig:4a}, the vertical axis represents the fraction of the true vacuum. In figure~\ref{fig:4b},  $D(t)$ and $C_{o}(t)$ represent the decoherence function and coherence, respectively. The parameters are set as: $a_{11}=a_{12}=a_{21}=0$, $a_{22}=1$, $\alpha=1$, $\beta=1$, $\gamma=1$, $\delta=500$.}
\label{fig:4}
\end{figure}

\begin{figure}[tbp]
\centering
\subfigure[]{
\begin{minipage}[t]{0.48\textwidth}
\centering
\includegraphics[width=7.5cm]{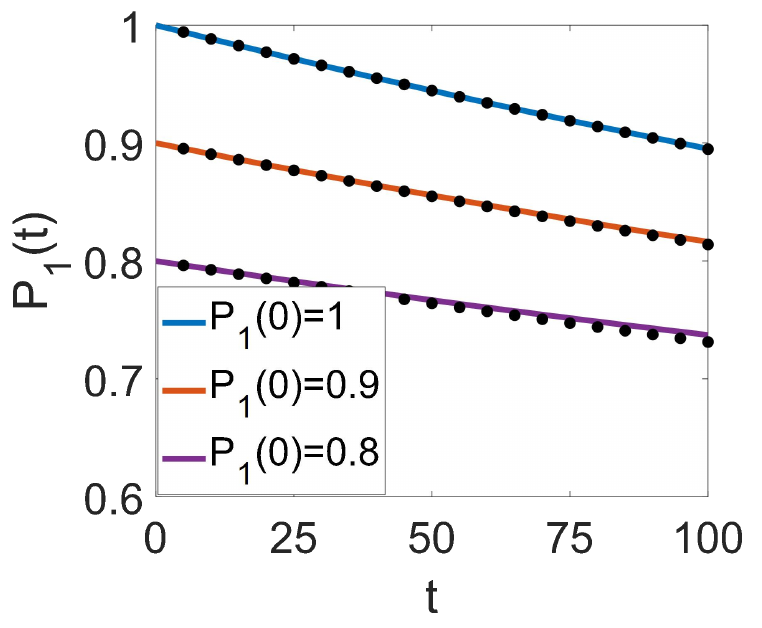}
\label{fig:5a}
\end{minipage}}
\subfigure[]{
\begin{minipage}[t]{0.48\textwidth}
\centering
\includegraphics[width=7cm]{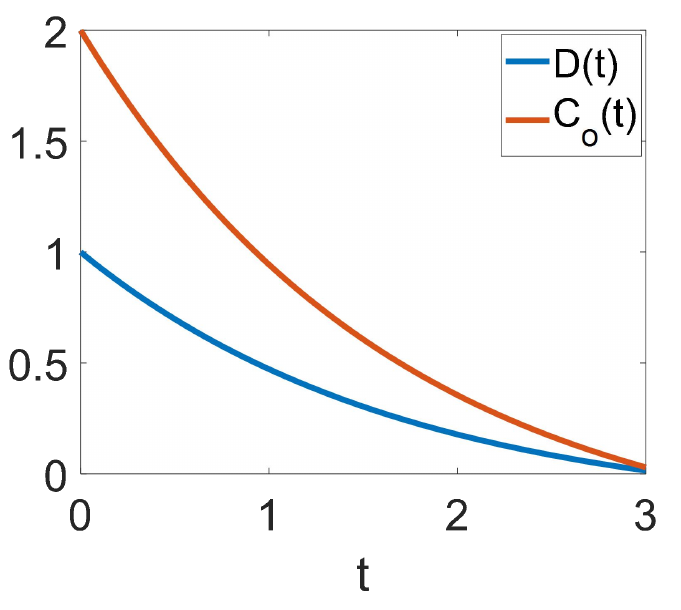}
\label{fig:5b}
\end{minipage}}
\caption{The evolution of certain quantities of the vacuum system. The horizontal axis of these two figures are the time variable. In figure~\ref{fig:5a}, the vertical axis represents the fraction of the true vacuum. The solid lines correspond to the quantum master equation, and the dotted lines correspond to the classical master equation. In figure~\ref{fig:5b},  $D(t)$ and $C_{o}(t)$ represent the decoherence function and coherence, respectively. The parameters are set as: $a_{11}=e^{2S_{E}(\Lambda_{1})+2S_{E}(\Lambda_{2})}= a_{12}=a_{21}=a_{22}$, $\alpha=1$, $\beta=1$. In figures~\ref{fig:5a},  $\gamma=50$ and $\delta=300$ . In figure~\ref{fig:5b}, $\gamma=0.3$ and $\delta=500$. }
\label{fig:5}
\end{figure}

Figures~\ref{fig:3}, ~\ref{fig:4} and~\ref{fig:5} show the evolution of some quantities of the system in the Markovian limit. In figure~\ref{fig:3}, we set  $a_{12}=a_{21}=a_{22}=0$. In this case, the quantum master equation \eqref{eq:3.18} becomes \eqref{eq:3.45}. Noted that the transition rate $\Gamma_{ij}$ in the classical master equation \eqref{eq:1.1} is~\cite{SF,ALI,JG}
\begin{equation}
\label{eq:4.8}
\Gamma_{ij}=A_{j}e^{-2S_{cl}-\mathbf{S}(\Lambda_{j})}.
\end{equation}
In equation \eqref{eq:4.8},  $2S_{cl}$ is the Euclidean action of the bounce solution, while $\mathbf{S}(\Lambda_{j})$ refers to the thermal entropy of the dS vacuum $\Lambda_{j}$. In some references~\cite{JGA,JG}, the prefactor $A_{j}$ is chosen as $(4\pi/3) H_{j}^{-3}$. In~\cite{ALI}, Linde set $A_{j}=1$. The prefactor is not important for this work (its effect can be included by modify the entropy). Thus, we also set $A_{j}=1$.

Combining equations \eqref{eq:3.46}, \eqref{eq:3.48} and \eqref{eq:4.8}, one can show that $a_{11}=e^{2S_{E}(\Lambda_{1})+2S_{E}(\Lambda_{2})}$. In figure~\ref{fig:4}, we set  $a_{11}=a_{12}=a_{21}=0$ and $a_{22}=1$. In this case, the quantum master equation \eqref{eq:3.18} becomes
\begin{equation}
\label{eq:4.9}
\frac{d\rho}{dt}=V\rho V^{\dag}-\frac{1}{2}\{V^{\dag}V,\rho\}.
\end{equation}
And in figure~\ref{fig:5}, we set $a_{11}=e^{2S_{E}(\Lambda_{1})+2S_{E}(\Lambda_{2})}=a_{12}=a_{21}=a_{22}$. Then equation \eqref{eq:3.18} becomes
\begin{equation}
\label{eq:4.10}
\frac{d\rho}{dt}=a_{11}\Big((K+V)\rho (K+V)^{\dag}-\frac{1}{2}\{(K+V)^{\dag}(K+V),\rho\}\Big).
\end{equation}

The initial state of the system is $\rho(0)$. For convenience, we rewrite $\rho(0)$ into the vector form
\begin{equation}
\label{eq:4.11}
u(0)\equiv(\rho_{11}(0), \rho_{12}(0), \rho_{21}(0), \rho_{22}(0)).
\end{equation}
Figure~\ref{fig:3a} shows the evolution of the comoving volume fraction of the true vacuum (the fraction of the false vacuum is equal to one minus the fraction of the true vacuum). In this figure, the solid lines are obtained by solving the quantum master equation \eqref{eq:3.18}, yet the dotted lines were obtained by solving the classical master equation \eqref{eq:1.1}. The red solid line corresponds to the initial state $u(0)=(1, 0, 0, 0)$. The blue solid line corresponds to the initial state $u(0)=(0.9, 0, 0, 0.1)$. And the green solid line corresponds to the initial state $u(0)=(0.8, 0, 0, 0.2)$. Figure~\ref{fig:3a} shows that the evolution of the fraction determined by the quantum master equation \eqref{eq:3.18} is consistent with the classical master equation \eqref{eq:1.1}.

Bringing  $a_{12}=a_{21}=a_{22}=0$ into the quantum master equation \eqref{eq:3.18}, one can easily prove that in the steady state, the fraction of the dS vacuum $\Lambda_{i}$ is
\begin{equation}
\label{eq:4.12}
P_{i}\propto e^{\mathbf{S}(\Lambda_{i})}.
\end{equation}
$\mathbf{S}(\Lambda_{i})$ is the thermal entropy of the dS vacuum. Equation \eqref{eq:4.12} is also the steady state solution of the classical master equation \eqref{eq:1.1}. This result is consistent with the Hartle-Hawking wave function~\cite{JH}. Equation \eqref{eq:4.12} shows that a larger cosmological constant corresponds to a lower fraction. Thus, in the steady state, the dominant vacuum corresponds to the smallest cosmological constant.

Figure~\ref{fig:3b} shows the variation in the comoving volume fraction of the vacua. $P_{1}(t)$ is the fraction of the true vacuum and $P_{2}(t)$ is the fraction of the false vacuum.
Different colors represent different values of the fraction. The upper half of this figure corresponds to the initial state $u(0)=(1, 0, 0, 0)$, while the lower half corresponds to the initial state $u(0)=(0, 0, 0, 1)$. This figure clearly shows that the false vacuum and the true vacuum can tunnel to one another. The curved spacetime promotes the true vacuum tunneling to the false vacuum.

Figure~\ref{fig:3c} shows the evolution of the coherence and the decoherence function. The coherence is defined as~\cite{TM}
\begin{equation}
\label{eq:4.13}
C_{o}(t)\equiv\sum_{i\neq j}|\rho_{ij}(t)|.
\end{equation}
Decoherence is the process of decreasing coherence. The system is equivalent to a two level system. Thus one can also introduce the decoherence function~\cite{HE}
\begin{equation}
\label{eq:4.14}
D(t)\equiv \frac{\rho_{12}(t)}{\rho_{12}(0)}
\end{equation}
to describe the decoherence process.  In figure~\ref{fig:3c}, the initial state is chosen as $u(0)=(1,1,1,0)$. The coherence of the initial state is not zero. One can see from figure~\ref{fig:3c} that both the coherence and the decoherence function decrease monotonically over time and ultimately tend towards zero. This indicates that decoherence has occurred and the initial quantum state will decohere into the final classical state.

The issue of decoherence of the universe has not been completely resolved yet, although significant numbers of the  studies have been performed~\cite{AOB,JJH,RJ,SH,CK1,YHW1,YHW2,BL1,BL2,BL3}. In~\cite{AOB,JJH,RJ,SH,CK1} , Kiefer, Barvinsky, Halliwell, Laflamme, and other researchers showed that environmental fluctuations lead to the decoherence of the inflationary universe. In~\cite{BL2}, Calzettaa, Hu and Mazzitellic derived the quantum master equation of the scalar field under the single-mode approximation. They showed that the expansion of the dS spacetime may lead to the decoherence of the long wave modes of the scalar field.  In~\cite{YHW1}, Yu showed that the quantum state of a freely falling two-level detector in dS spacetime can decohere to a classical state. Our results show that the spacetime leads to the  decoherence of the vacuum state of the scalar field. This may help to understand the decoherence of the universe  from another angle.

Figure~\ref{fig:3c} indicates that in the final state, the non-diagonal elements of the reduced density matrix are equal to zero. And figure~\ref{fig:3a} shows that the evolution of the diagonal elements  is in agreement with the results obtained from the classical master equation. Therefore, the classical limit of the quantum master equation \eqref{eq:3.18} is consistent with the classical master equation \eqref{eq:1.1}. The quantum master equation \eqref{eq:3.18} is capable of characterizing the decoherence process of the vacuum system. This may help to understand the decoherence of the quantum universe.

Figure~\ref{fig:3d} shows the flux variation. The red solid line corresponds to the initial state $u(0)=(1, 0, 0, 0)$. The blue solid line corresponds to the initial state $u(0)=(0.9, 0, 0, 0.1)$. And the green solid line corresponds to the initial state $u(0)=(0.8, 0, 0, 0.2)$. Thus, in the initial state, the fraction of the true vacuum state is larger than the false vacuum state. The flux between the true vacuum state and the false vacuum state is defined as~\cite{JW,JW2}
\begin{equation}
\label{eq:4.15}
F_{12}(t)\equiv\Gamma_{12}P_{2}(t)-\Gamma_{21}P_{1}(t).
\end{equation}
From figure~\ref{fig:3d}, one can see that the value of $F_{12}(t)$ is negative. This indicates that the fraction of the true vacuum state decreases with time. As the flux approaches zero, the distribution of the vacuum state reaches a steady-state, wherein detailed balance is preserved. However, in the non-steady state, the flux is non-zero, indicating that the evolution is irreversible.


Figure~\ref{fig:4} corresponds to the master equation \eqref{eq:4.9}. Figure~\ref{fig:4a} shows the variation of the fraction of the true vacuum. Figure~\ref{fig:4b} shows the variation of the coherence and decoherence function. In figure~\ref{fig:4a}, the red solid line corresponds to the initial state $u(0)=(0.9, 0, 0, 0.1)$. The blue solid line corresponds to the initial state $u(0)=(0.8, 0, 0, 0.2)$. And the green solid line corresponds to the initial state $u(0)=(0.7, 0, 0, 0.3)$. In figure~\ref{fig:4b}, the initial state is chosen as $u(0)=(1, 1, 1, 0)$. Figure~\ref{fig:4a} shows that if $a_{11}=a_{12}=a_{21}=0$, then the fraction of the vacuum state does not change with time. This indicates that the vacuum decay does not occur. However, figure~\ref{fig:4b} shows that both the coherence and decoherence function continuously decrease over time. Thus in this case, although the vacuum decay does not occur, the decoherence still takes place.

Figure~\ref{fig:5} corresponds to equation \eqref{eq:4.10}. In figure~\ref{fig:5a}, the blue solid line corresponds to the initial state $u(0)=(1, 0, 0, 0)$. The brown solid line corresponds to the initial state $u(0)=(0.9, 0, 0, 0.1)$. And the purple solid line corresponds to the initial state $u(0)=(0.8, 0, 0, 0.2)$.  The solid lines are obtained by simulating equation \eqref{eq:4.10}. And the dotted lines are obtained by simulating the classical master equation \eqref{eq:1.1}. In figure~\ref{fig:5b}, the initial state is chosen as $u(0)=(1, 1, 1, 0)$. Figure~\ref{fig:5a} shows that the quantum master equation is consistent with the classical master equation in this case. Figure~\ref{fig:5b} shows that the decoherence has taken place, resulting in a final state that is classical.

To sum up, figures~\ref{fig:3}, ~\ref{fig:4} and~\ref{fig:5} correspond to the Markovian limit where both the coherence and decoherence function monotonically decrease over time. These results show the decoherence of the quantum vacuum system into a classical system.  In some cases (such as figure~\ref{fig:3} and figure~\ref{fig:5}), the classical limit of the quantum master equation \eqref{eq:3.18} is consistent with the classical master equation. These simulations were performed for the two dS vacua system, but we believe that the results would not change qualitatively when increasing the number of dS vacua. Despite introducing the Markovian approximation in equation \eqref{eq:3.25}, it does not imply that the quantum master equation \eqref{eq:3.18} is a Markovian master equation. Equations \eqref{eq:3.29}-\eqref{eq:3.33} show that the coefficients $a_{ij}$ may depend on the time variable, which suggests that the quantum master equation \eqref{eq:3.18} could have non-Markovian properties in certain situations. In 2008, Winitzki also showed that vacuum decay dynamics may entail non-Markovian
correlations~\cite{SW}.

\section{Conclusions and discussions}
\label{sec:5}
There is no time variable in the Wheeler-DeWitt equation. This induced the time problem in quantum gravity. However, our interest lies in the dynamical information of the subsystem of the universe.  The subsystem is not an isolated system, usually the general covariance does not constrain that the (effective) Hamiltonian of the subsystem must be zero. Thus, maybe one can use the von Neumann equation \eqref{eq:2.6} to describe the quantum dynamics of the subsystem. Equation \eqref{eq:2.6} can also be derived from the Wheeler-DeWitt equation by introducing the Brown-Kucha$\check{\mathrm{r}}$ dust field. Therefore, the time variable in equation \eqref{eq:2.6} can also be interpreted as the dust field. Although the generalized covariant principle leads to the difficulty to obtain all quantum dynamical information of the universe, the von Neumann equation \eqref{eq:2.6} indicates that we can still obtain partial quantum dynamical information of the universe. The von Neumann equation \eqref{eq:2.6} serves as the fundamental equation in our work.

Starting from the von Neumann equation \eqref{eq:2.6}, after tracing out the environment, we obtained the quantum master equation \eqref{eq:3.18} by two different methods. In the derivation process, we did not introduce the secular approximation. Thus equation \eqref{eq:3.18} can also be referred to as the Redfield equation. We also did not introduce the Born approximation since the interaction Hamiltonian cannot be viewed as a small value. Thus, the coefficients $a_{ij}$ may change over time.  Therefore, equation \eqref{eq:3.18} may exhibit non-Markovian properties in some situations. Equation \eqref{eq:3.18} can be used to describe the evolution of the comoving volume fraction of the vacua, as well as certain quantum quantities of the vacuum system such as coherence.  Equation \eqref{eq:3.18} can also describe the evolution of the superposition state of the vacuum. However, the classical master equation \eqref{eq:1.1} can only describe the evolution of the comoving volume fraction of the vacuum.

The entropy of the dS spacetime is not zero, which indicates that the dS spacetime has micro degrees of freedom. We are not interested in the microstates. Thus we need to coarse-grain the physical Hilbert space of the system. Coarse graining contributes a degree of degeneracy to the vacuum state and destroys the Hermiticity of the kinetic energy operator. Similarly, it can also destroy the Hermiticity of the Hamiltonian operator. Therefore, if we are interested in the coarse-grained dynamical information, the evolution of the coarse-grained system may be non-unitary even for an isolated system. In other words, coarse grained information is not conserved.

Finally, we simulated the quantum master equation \eqref{eq:3.18} in the Markovian limit. We found that in some cases, the classical limit of the quantum master equation is consistent with the classical master equation. We show that both the coherence and decoherence function decrease monotonically with time. This indicates that the decoherence has emerged.  Consequently, the initial quantum state will decohere to the final classical state. This helps the understanding of the decoherence of the universe. We also show that the absolute value of the flux decreases with time and eventually approaches zero, indicating that the evolution in time is irreversible and in the steady state, detailed balance is maintained. Therefore, in the steady state, different vacua are in equilibrium, and the dominant vacuum corresponds to the smallest cosmological constant.

\section*{Acknowledgements}
Hong Wang  was supported by the National Natural Science Foundation of China Grant No. 21721003.  No.12234019. Hong Wang thanks for the help from Professor Erkang Wang.

\appendix

\end{document}